\documentclass[aps,prd,10pt,notitlepage,nofootinbib]{revtex4-1}

\usepackage[utf8]{inputenc}
\usepackage{amsmath,amssymb,amsfonts}
\usepackage{graphicx, overpic}
\usepackage{palatino}

\usepackage{bbold}
\usepackage{bm}
\usepackage[usenames,dvipsnames]{xcolor}
\usepackage{color}
\usepackage[colorlinks=true,linkcolor=blue,urlcolor=blue,citecolor=blue]{hyperref}

\usepackage{slashed}
\usepackage[english]{babel}
\usepackage{dcolumn}
\usepackage{pifont}
\usepackage{dsfont,mathrsfs}
\usepackage{cancel}
\usepackage{bigints}
\usepackage{accents}
\usepackage{soul}
\usepackage{multirow}
\usepackage{natbib}
\usepackage{tikz-feynman}

\usepackage{amsmath, amssymb}
\usepackage{bbold}
\usepackage{makecell}
\usepackage{simpler-wick}
\usepackage[export]{adjustbox}

\usepackage[bb=boondox]{mathalpha}

\usepackage{tikz}
\usepackage{tikz-feynman}
\tikzfeynmanset{compat=1.1.0} 

\usepackage{orcidlink} 

\newcommand{\nn}{\nonumber}

\newcommand{\MB}[1]{\left|#1\right|}
\newcommand{\FB}[1]{\left(#1\right)}
\newcommand{\SB}[1]{\left\{#1\right\}}
\newcommand{\TB}[1]{\left[#1\right]}

\newcommand{\munu}{{\mu\nu}}

\newcommand{\IM}{\text{Im}}

\newcommand{\half}{\dfrac{1}{2}}
\newcommand{\gm}{\gamma}

\newcommand{\Tr}[1]{{\rm Tr}\TB{#1}}

\newcommand{\ov}[1]{\overline{#1}}

\newcommand{\mO}{\mathcal{O}}

\newcommand{\mM}{\mathcal{M}}

\allowdisplaybreaks

\begin{document}
	\title{High energy thermal photons from chirally imbalanced QGP}

	\author{Sourav Duari\orcidlink{0009-0006-0795-5186}$^{a,c}$}
	\email{s.duari@vecc.gov.in}
	\email{sduari.vecc@gmail.com}
	
	\author{Nilanjan Chaudhuri\orcidlink{0000-0002-7776-3503}$^{a}$}
	\email{n.chaudhri@vecc.gov.in}
	\email{nilanjan.vecc@gmail.com}

	\author{Pradip Roy\orcidlink{0009-0002-7233-4408}$^{b,c}$}
	\email{pradipk.roy@saha.ac.in}	
	
	\author{Sourav Sarkar\orcidlink{0000-0002-2952-3767}$^{a,c}$}
	\email{sourav.vecc@gmail.com}

	\affiliation{$^a$Variable Energy Cyclotron Centre, 1/AF Bidhannagar, Kolkata - 700064, India}
	\affiliation{$^b$Saha Institute of Nuclear Physics, 1/AF Bidhannagar, Kolkata - 700064, India}
	\affiliation{$^c$Homi Bhabha National Institute, Training School Complex, Anushaktinagar, Mumbai - 400085, India}
%
	
	\begin{abstract}
		We compute the thermal photon production rate from a chirally asymmetric quark gluon plasma. We estimate the hard part from scattering of chiral quarks and demonstrate the cut-off independence of the total thermal photon emission rate. This is achieved by combining the hard with the soft contribution evaluated in our previous work~\cite{Chaudhuri:2025kpl} using the Braaten-Pisarski method of hard thermal loop perturbation theory. It is observed that the presence of chiral imbalance leads to an overall enhancement in the emission rate of thermal photons.

	\end{abstract}
	
	\maketitle
		\section{Introduction}
\label{Sec_Intro}
 The nontrivial topological structure of the vacuum is a fundamental feature of non-Abelian gauge theories underlying the strong and electroweak sectors of the Standard Model of particle physics. The compact nature of non-Abelian gauge groups permits topologically nontrivial gauge field configurations, giving rise to a vacuum structure composed of an infinite set of degenerate but topologically distinct sectors separated by potential barriers~\cite{Jackiw:1976pf,Callan:1976je}. These vacuum sectors are characterized by different winding (or Chern--Simon) numbers and are connected through transitions mediated by gauge field configurations of nontrivial topology~\cite{Shifman:1988zk,Lenz:2001me}. At zero temperature, transitions between distinct vacua occur via quantum tunneling processes induced by instantons~\cite{Belavin:1975fg,tHooft:1976rip,tHooft:1976snw}. At finite temperature, however, thermally activated transitions over the potential barrier become possible through sphaleron processes, whose abundance is expected to increase substantially in hot non-Abelian matter~\cite{Kuzmin:1985mm,Arnold:1987mh,Khlebnikov:1988sr,Arnold:1987zg}. 
 
 In the electroweak sector, the height of the potential barrier separating topological vacuum sectors is determined by the Higgs condensate and is generally too large to allow an experimental observation of vacuum transitions in present collider environments. In contrast, Quantum Chromodynamics (QCD) possesses analogous topological vacuum sectors separated by potential barriers of order $\Lambda_{\rm QCD}$, making topological transitions phenomenologically relevant in hot strongly interacting matter. Relativistic heavy-ion collisions provide a unique environment to study such effects by creating a rapidly evolving, strongly interacting quark--gluon plasma (QGP) at temperatures where sphaleron-induced transitions are expected to occur at an appreciable rate. Topological gauge field configurations in QCD are believed to play an important role in nonperturbative phenomena such as spontaneous chiral symmetry breaking and possibly confinement. Furthermore, owing to the axial anomaly, these topological transitions violate axial charge conservation and generate an imbalance between right- and left-handed quarks~\cite{Adler:1969gk,Bell:1969ts}. This chirality imbalance is commonly described by introducing a chiral chemical potential, $\mu_5$, which parametrizes the difference between right- and left-handed quark number densities.  The dynamics of massless chiral fermions has attracted considerable attention over the past two decades in a variety of physical settings, including hot and dense QCD matter realized in the QGP~\cite{Son:2012zy,Akamatsu:2013pjd,Duari:2025kar}, early-Universe cosmology, and cold dense matter in compact astrophysical objects such as neutron stars~\cite{Kaplan:2016drz,Grabowska:2014efa,Yamamoto:2015gzz,Huang:2017pqe,Sen:2016jzl}.

 Although QCD does not exhibit a global violation of $CP$ symmetry, local $P$- and $CP$-odd domains may arise dynamically due to fluctuations of topological charge in hot QCD matter~\cite{McLerran:1990de,Moore:2010jd}. Consequently, individual heavy-ion collision events may develop a net chirality of either sign, corresponding to an excess of right- or left-handed quarks generated through topological vacuum transitions. One of the most prominent proposed signatures of such chirality imbalance is the chiral magnetic effect (CME), wherein an imbalance of chirality, in the presence of a strong magnetic field generated by the colliding ions, induces an electric current along the direction of the magnetic field~\cite{Kharzeev:2007jp,Fukushima:2008xe,Kharzeev:2012ph,Kharzeev:2015znc,Manuel:2015zpa,Koch:2016pzl}. Extensive experimental searches for the CME have been carried out through azimuthal particle correlation measurements in relativistic heavy-ion collisions. However, no conclusive evidence has yet been established due to substantial background contributions to the relevant observables (see~\cite{Li:2025yxx} for a recent review). In addition, anomaly-induced transport phenomena have been extensively investigated in condensed matter systems, particularly in Dirac and Weyl semimetals, where chiral quasiparticles emerge as low-energy excitations~\cite{Son:2012bg,Gorbar:2013dha,Li:2014bha,Cortijo:2016wnf,Kaushik:2018tjj,Sukhachov:2021fkh,Jamal:2026rjk}.

 Given the difficulty of establishing unambiguous signatures of chirality imbalance through hadronic observables alone, electromagnetic (EM) probes such as photons and dileptons offer a complementary and potentially cleaner avenue to investigate anomalous phenomena in hot QCD matter. Owing to their large mean free path relative to the characteristic size of the medium, photons and dileptons escape the strongly interacting plasma with minimal final-state interactions and therefore retain information about the thermodynamic and dynamical properties of the system at the time of their production. Consequently, EM observables have long been regarded as sensitive probes of the early-time evolution of the QGP produced in relativistic heavy-ion collisions at RHIC and LHC energies. In particular, it has been proposed that electromagnetic emissions may provide indirect signatures of chirality imbalance. For example, the dilepton production rate (DPR) in a hot and dense chiral medium have been investigated in Ref.~\cite{Chaudhuri:2022rwo}, where it is shown that the DPR is significantly enhanced in the low invariant-mass region due to Landau-cut contributions, resulting in the disappearance of the forbidden gap between the Landau and Unitary cuts at sufficiently high temperature and chiral chemical potential. Photon production mechanisms stemming from anomalous responses in abelian chiral plasmas have recently been investigated under various configurations, focusing on radiative instabilities and time-dependent chiral chemical potentials~\cite{Tuchin:2019jxd,Tuchin:2025stl,Tuchin:2025bll,Kroth:2026ykh}. Furthermore, spin-polarized photon and dilepton emission from a chiral plasma has been conjectured to provide possible signatures of local $P$- and $CP$-odd effects in strong interactions~\cite{Mamo:2013jda}. In a different context, a novel mechanism for soft photon production based on the interplay between the conformal anomaly of QCD and QED and strong electromagnetic fields generated in heavy-ion collisions has also been proposed, yielding substantial contributions to both the azimuthal anisotropy and radial flow of direct photons~\cite{Basar:2012bp}.

 In this work, we extend our previous study~\cite{Chaudhuri:2025kpl}, in which the soft contribution to the thermal photon emission rate from a chirally imbalanced QGP has been investigated. The present analysis focuses on the hard contribution to the real photon production rate obtained by evaluating the scattering matrix elements for quark--antiquark annihilation and QCD Compton scattering processes within first-order perturbation theory. Owing to the presence of chiral imbalance, the contributions from left- and right-handed quarks are treated separately. To regulate the infrared divergence of the invariant amplitudes, an intermediate momentum cutoff parameter, $k_c$, is introduced to separate hard and soft momentum scales. The soft contribution is incorporated using the Braaten--Pisarski resummation framework and is adopted from our previous work~\cite{Chaudhuri:2025kpl}. Particular emphasis is placed on establishing the cutoff independence of the total photon emission rate through a detailed analysis of the individual hard and soft contributions. While the cancellation of the intermediate cutoff dependence at finite baryon density has previously been demonstrated in Ref.~\cite{Traxler:1994hy}, the present work generalizes this framework to a QGP with finite chiral imbalance.

 The plan of the paper is as follows. In Sec.~\ref{Sec_softpart} we have noted down the results for soft contribution from~\cite{Chaudhuri:2025kpl} followed by calculation for the hard part in Sec.~\ref{Sec_hardpart}. The cut-off independence of the total photon emission rate is discussed in Sec.~\ref{Sec_cutoff}. In Sec.~\ref{Sec_result} we show the numerical results followed by a brief summary in Sec.~\ref{Sec_summary}.

\section{Setup}

The emission rate of real photons with energy $E$ and momentum $\vec{p}$ from a thermalized  plasma of quarks and gluons is expressed as~\cite{Kapusta:1991qp,Baier:1991em,Kapusta:2006pm,Bellac:2011kqa,Haque:2024gva}
\begin{equation}
	E \frac{dR}{d^3 p} = \frac{2}{(2\pi)^3} ~\IM~ ^{\rm ret}\Pi^\mu_\mu (E,\vec{p}) ~\frac{1}{e^{\beta E} -1} \label{Eq_Rate}
\end{equation}
where $^{\rm ret}\Pi^\mu_\mu$ is the retarded self-energy in a thermal medium. It is to be noted that this formula is exact to all orders in strong interaction and only up to leading order in electromagnetic interaction due to the assumption that the produced photon emerges without final state scattering from the hot matter~\cite{Kapusta:1991qp,Kapusta:2006pm}. 
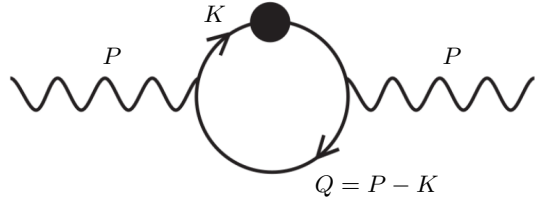
\begin{figure}[h]
	\begin{overpic}[scale=0.26]{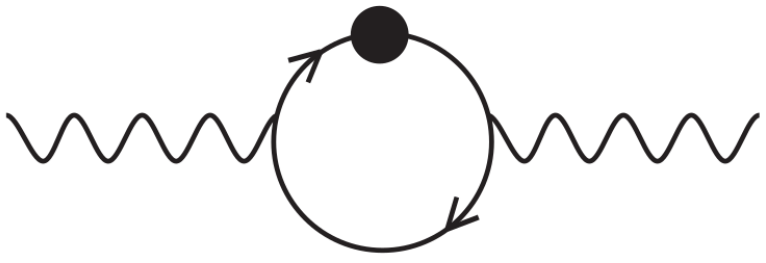}
		\put(58,-2){$Q=P-K$}
		\put(37,30){$K$}
		\put(18,22){$P$}
		\put(82,22){$P$}
	\end{overpic}
	\caption{The photon self-energy is evaluated in the HTL approximation. The internal fermion line with a black blob represents the soft quark propagator dressed with HTL resummation, while the line without the blob denotes the bare hard quark propagator. }
	\label{Fig_photon}
\end{figure}
It is well-known that tree-level calculations of thermal processes involving the exchange of massless particles give rise to infrared divergences which are screened by thermal loop corrections~\cite{Braaten:1991dd}. This in turn results in decomposition of hard photon production rate into two parts.  The hard part, that is when the momenta of exchanged quark is large, is obtained in first order perturbation theory using bare propagators. On the other hand, the soft contribution is calculated employing the resummed propagators for intermediate quarks with soft momenta. Assuming the weak coupling limit, an intermediate scale $k_c$ is introduced  demanding $gT \ll k_c \ll T$ which separates different momentum scales of the exchanged quark. Here $T$ is the system temperature and $g$ is the strong coupling constant. The final results is independent of separation scale $k_c$~\cite{Kapusta:1991qp, Baier:1991em, Kapusta:2006pm, Bellac:2011kqa}.

\subsection{Contribution from soft momentum transfer}
\label{Sec_softpart}

The soft contribution to the emission rate of real photons from quark gluon plasma with chiral imbalance has been recently obtained in~\cite{Chaudhuri:2025kpl} from the imaginary part of the retarded photon self-energy as depicted in Fig.~\ref{Fig_photon}. Since the energy of the produced photon is high, due to kinematics there can only be one full quark propagator in the polarization tensor, since the other one has to be hard. Moreover no vertex correction is necessary as the high energy photon resolves the vertex completely. The final expression for the soft contribution to the thermal photon production rate with energy $E$ is given by
\begin{equation}
	2E\frac{dR^{\rm \ soft}}{d^3 p} = \dfrac{5 \alpha }{12 \pi^3} e^{-\beta E} \TB{ M_L^2 \ln \dfrac{k_c^2}{2 M_L^2} + M_R^2 \ln \dfrac{k_c^2}{2 M_R^2}}~\label{Eq_Soft_cont}
\end{equation}
with
\begin{equation}
		M_{L/R}^2 = \dfrac{g^2 C_F}{8}\TB{T^2 + \frac{\mu^2}{\pi^2} + \frac{\mu_5^2}{\pi^2}} \pm  \dfrac{g^2 C_F}{4\pi^2} \mu \mu_5 = \dfrac{g^2 C_F}{8} \TB{T^2 + \dfrac{\FB{\mu \pm \mu_5}^2}{\pi^2} } ~.
\end{equation}
where $\mu$ and $\mu_5$ are the quark and chiral chemical potentials respectively and $C_F = {4}/{3}$ is the Casimir invariant of the gauge group $SU(3)$ in the fundamental representation and $\alpha = e^2/4\pi$. Effective quark propagator in a presence of chiral asymmetry necessary for evaluating the above results is obtained in~\cite{Duari:2025iah}.  
While calculating the soft contribution as expressed by Eq.~\eqref{Eq_Soft_cont} we have assumed that the exchanged quark propagator ia dressed and satisfies a hyperbolic cut-off $0 \le k^2 - \omega^2 \le k_c^2$ in accordance with Refs.~\cite{Kapusta:1991qp,Baier:1991em}.  This cut-off will also be introduced in the calculation of the hard contribution to regulate the infrared divergences  arising from the soft region of phase-space in the $u$ and $t$-channel diagrams.
From Eq.~\eqref{Eq_Soft_cont} it can be seen that, as a consequence of non zero values of $\mu_5$, the soft part of the hard photon production rate receives different contributions from the left- and right-handed quarks present in the medium. This is owing to the fact that in a chirally imbalanced medium left- and right-handed quark follow non-degenerate dispersion relation and acquire different thermal mass denoted by $M_L$ and $M_R$ respectively~\cite{Duari:2025iah}. It should be noted that, in the limit $\mu_5 \to 0$, Eq.~\eqref{Eq_Soft_cont} correctly reduces to the result of Ref.~\cite{Traxler:1994hy} where hard photon rate from hot and dense QGP is calculated. This can be easily checked since $M_{R/L} \to g^2 C_F (T^2 +\mu^2/\pi^2) {/8} $ in absence of $\mu_5$ and $\alpha_S = g^2/4\pi$. Furthermore at $\mu \to 0$ we recover the well known expression given in Refs.~\cite{Kapusta:1991qp,Baier:1991em}.

\subsection{Contribution from hard momentum transfer}
\label{Sec_hardpart}

To calculate the hard part of the real photon rate one has to consider scattering matrix elements corresponding to quark-antiquark annihilation and QCD Compton scattering using first order perturbation theory~\cite{Kapusta:1991qp,Baier:1991em}. 
Here we consider scattering processes involving left- and right-handed  quarks separately due to the presence of chiral imbalance~\cite{Peskin:1995ev}. 
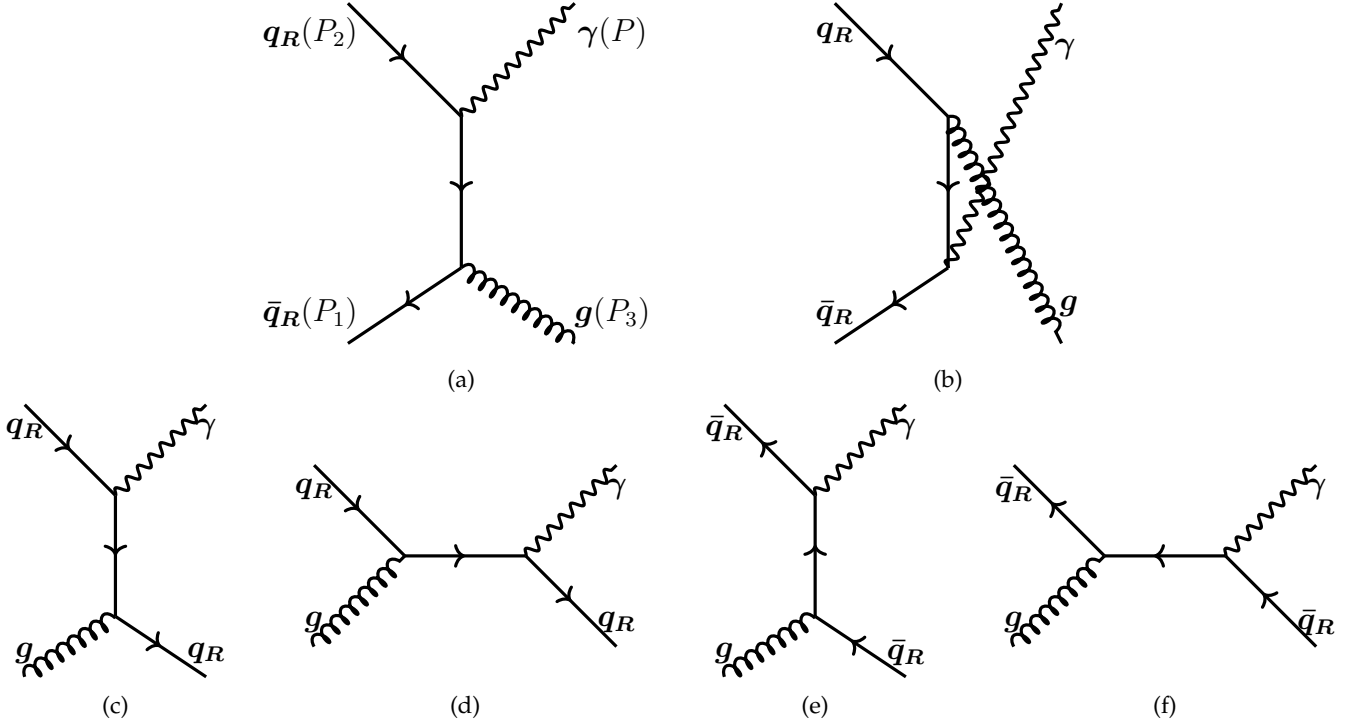
\begin{figure}[h]
	\centering
	
	\begin{tikzpicture}[scale=1.]
		
		\coordinate (v1) at (0,0);
		\coordinate (v2) at (0.0,-2.0);
		
		
		\draw[line width=1.2pt,
		postaction={decorate},
		decoration={
			markings,
			mark=at position 0.5 with {\arrow{>}}
		}
		] (-1.5,1.5) -- (v1)
		node at (-2.0,1.1) {\large $\bm{q_R} (P_2)$};

		\draw[line width=1.2pt,decorate, decoration={snake, amplitude=2pt, segment length=6pt}]
		(v1) -- (1.5,1.5)
		node at (2.0,1.1)  {\large $\bm{\gamma} (P)$};
		
		\draw[line width=1.2pt,
		postaction={decorate},
		decoration={
			markings,
			mark=at position 0.5 with {\arrow{>}}
		}
		] (v1) -- (v2);
		
	\draw[line width=1.2pt,
	postaction={decorate},
	decoration={
		markings,
		mark=at position 0.5 with {\arrow{>}}
	}
	] (v2) -- (-1.5,-3)
	
	node at (-2.0,-2.6) {\large $\bm{{\bar{q}_R}} (P_1)$};
		
		\draw[line width=1.2pt,decorate, decoration={coil, amplitude=3pt, segment length=5pt}]
		(v2) -- (1.5,-3)
		node at (2.0, -2.6) {\large $ \bm{g} (P_3) $};
		
		\node at (0,-3.5) {(a)};
		
		
	\end{tikzpicture}
	\qquad\qquad\qquad
		\begin{tikzpicture}[scale=1.]
		
		\coordinate (v1) at (0,0);
		\coordinate (v2) at (0.0,-2.0);
		
		
		\draw[line width=1.2pt,
		postaction={decorate},
		decoration={
			markings,
			mark=at position 0.5 with {\arrow{>}}
		}
		] (-1.5,1.5) -- (v1)
		node at (-1.5,1.1) {\large $\bm{q_R}$};

		\draw[line width=1.2pt,decorate, decoration={snake, amplitude=2pt, segment length=6pt}]
		(v2) -- (1.5,1.5)
		node at (1.6,-2.5)  {\large $\bm{g}$};
		
		\draw[line width=1.2pt,
		postaction={decorate},
		decoration={
			markings,
			mark=at position 0.5 with {\arrow{>}}
		}
		] (v1) -- (v2);
		
		\draw[line width=1.2pt,
		postaction={decorate},
		decoration={
			markings,
			mark=at position 0.5 with {\arrow{>}}
		}
		] (v2) -- (-1.5,-3)
		
		node at (-1.5,-2.6) {\large $\bm{{\bar{q}_R}}$};
		
		\draw[line width=1.2pt,decorate, decoration={coil, amplitude=3pt, segment length=5pt}]
		(v1) -- (1.5,-3)
		node at (1.55, 0.9) {\large $ \bm{\gm} $};
		
		\node at (0,-3.5) {(b)};
		
		
	\end{tikzpicture}

	\begin{tikzpicture}[scale=0.8]
		
		\coordinate (v1) at (0,0);
		\coordinate (v2) at (0,-2);
		
		\draw[line width=1.2pt,
		postaction={decorate},
		decoration={markings, mark=at position 0.5 with {\arrow{>}}}
		] (-1.5,1.5) -- (v1)
		node at (-1.5,1.1) {\large $\bm{q_R}$};
		
		\draw[line width=1.2pt, decorate,
		decoration={coil, amplitude=3pt, segment length=5pt}
		] (-1.5,-3) -- (v2)
		node at (-1.5,-2.6) {\large $\bm{g}$};
		
		\draw[line width=1.2pt,
		postaction={decorate},
		decoration={markings, mark=at position 0.5 with {\arrow{>}}}
		] (v1) -- (v2);
		
		\draw[line width=1.2pt, decorate,
		decoration={snake, amplitude=2pt, segment length=6pt}
		] (v1) -- (1.5,1.5)
		node at (1.5,1.1) {\large $\bm{\gamma}$};
		
		\draw[line width=1.2pt,
		postaction={decorate},
		decoration={markings, mark=at position 0.5 with {\arrow{>}}}
		] (v2) -- (1.5,-3)
		node at (1.5,-2.6) {\large $\bm{q_R}$};
		
		\node at (0,-3.5) {(c)};
		
	\end{tikzpicture}
	\qquad
	\begin{tikzpicture}[scale=0.8]
		
		\coordinate (v1) at (0,0);
		\coordinate (v2) at (2,0);
		
		\draw[line width=1.2pt,
		postaction={decorate},
		decoration={markings, mark=at position 0.5 with {\arrow{>}}}
		] (-1.5,1.5) -- (v1)
		node at (-1.5,1.1) {\large $\bm{q_R}$};
		
		\draw[line width=1.2pt, decorate,
		decoration={coil, amplitude=3pt, segment length=5pt}
		] (-1.5,-1.5) -- (v1)
		node at (-1.5,-1.1) {\large $\bm{g}$};
		
		\draw[line width=1.2pt,
		postaction={decorate},
		decoration={markings, mark=at position 0.5 with {\arrow{>}}}
		] (v1) -- (v2);
		
		\draw[line width=1.2pt, decorate,
		decoration={snake, amplitude=2pt, segment length=6pt}
		] (v2) -- (3.5,1.5)
		node at (3.5,1.1) {\large $\bm{\gamma}$};
		
		\draw[line width=1.2pt,
		postaction={decorate},
		decoration={markings, mark=at position 0.5 with {\arrow{>}}}
		] (v2) -- (3.5,-1.5)
		node at (3.5,-1.1) {\large $\bm{q_R}$};
		
		\node at (1,-2.5) {(d)};
		
	\end{tikzpicture}
	\qquad
	\begin{tikzpicture}[scale=0.8]
	
	\coordinate (v1) at (0,0);
	\coordinate (v2) at (0,-2);
	
	\draw[line width=1.2pt,
	postaction={decorate},
	decoration={markings, mark=at position 0.5 with {\arrow{<}}}
	] (-1.5,1.5) -- (v1)
	node at (-1.5,1.1) {\large $\bm{{\bar{q}_R}}$};
	
	\draw[line width=1.2pt, decorate,
	decoration={coil, amplitude=3pt, segment length=5pt}
	] (-1.5,-3) -- (v2)
	node at (-1.5,-2.6) {\large $\bm{g}$};
	
	\draw[line width=1.2pt,
	postaction={decorate},
	decoration={markings, mark=at position 0.5 with {\arrow{<}}}
	] (v1) -- (v2);
	
	\draw[line width=1.2pt, decorate,
	decoration={snake, amplitude=2pt, segment length=6pt}
	] (v1) -- (1.5,1.5)
	node at (1.5,1.1) {\large $\bm{\gamma}$};
	
	\draw[line width=1.2pt,
	postaction={decorate},
	decoration={markings, mark=at position 0.5 with {\arrow{<}}}
	] (v2) -- (1.5,-3)
	node at (1.5,-2.6) {\large $\bm{{\bar{q}_R}}$};
	
	\node at (0,-3.5) {(e)};
	
\end{tikzpicture}
\qquad
	\begin{tikzpicture}[scale=0.8]
	
	\coordinate (v1) at (0,0);
	\coordinate (v2) at (2,0);
	
	\draw[line width=1.2pt,
	postaction={decorate},
	decoration={markings, mark=at position 0.5 with {\arrow{<}}}
	] (-1.5,1.5) -- (v1)
	node at (-1.5,1.1) {\large $\bm{{\bar{q}_R}}$};
	
	\draw[line width=1.2pt, decorate,
	decoration={coil, amplitude=3pt, segment length=5pt}
	] (-1.5,-1.5) -- (v1)
	node at (-1.5,-1.1) {\large $\bm{g}$};
	
	\draw[line width=1.2pt,
	postaction={decorate},
	decoration={markings, mark=at position 0.5 with {\arrow{<}}}
	] (v1) -- (v2);
	
	\draw[line width=1.2pt, decorate,
	decoration={snake, amplitude=2pt, segment length=6pt}
	] (v2) -- (3.5,1.5)
	node at (3.5,1.1) {\large $\bm{\gamma}$};
	
	\draw[line width=1.2pt,
	postaction={decorate},
	decoration={markings, mark=at position 0.5 with {\arrow{<}}}
	] (v2) -- (3.5,-1.5)
	node at (3.5,-1.1) {\large $\bm{{\bar{q}_R}}$};
	
	\node at (1,-2.5) {(f)};
	
\end{tikzpicture}

	\caption{Diagrams (a) and (b) correspond to the annihilation of a right-handed quark and anti-quark. Diagrams (c)-(f) represent QCD Compton processes for right-handed quark and anti-quark respectively. Similar diagrams for left-handed quarks will also contribute. Momentum convention shown in diagram (a) is followed in all other diagrams.}
	\label{Fig_hard}
\end{figure}

 Let us first concentrate on the annihilation process for right-handed quarks as shown in Figs.~\ref{Fig_hard} (a) and (b). The invariant amplitude for the $t$ and $u$-channel diagrams shown in Figs.~\ref{Fig_hard} (a) and (b) respectively are given by 
\begin{align}
	\mM^{R,q\bar{q}\to g \gm }_t &= - e g e_q \SB{\bar{v}_R (P_1)\gm^\mu t^a \dfrac{\slashed{P}_2 -\slashed{P}}{t^2 +i \epsilon} \gm^\nu u_R (P_2)  } \epsilon_\mu (P_3) \epsilon_\nu (P) \\
	\mM^{R,q\bar{q}\to g \gm }_u &= -e g e_q  \SB{\bar{v}_R (P_1)\gm^\nu  \dfrac{\slashed{P}_2 -\slashed{P}_3}{u^2 +i \epsilon} \gm^\mu t^a u_R (P_2)  } \epsilon_\mu (P_3) \epsilon_\nu (P) 
\end{align}
Since there is fermion exchange the total amplitude is given by
\begin{align}
	\mM^{R,q\bar{q}\to g \gm }_{\rm tot} &= \mM^{R,q\bar{q}\to g \gm }_t+ \mM^{R,q\bar{q}\to g \gm }_u = e g e_q  {\bar{v}_R (P_1)t^a \Gamma_{R,q\bar{q}\to g \gm}^{\munu}u_R (P_2)  } \epsilon_\nu (P_3) \epsilon_\mu (P)
\end{align}
where $\Gamma_{R,q\bar{q}\to g \gm}^\munu =  \gm^\mu \dfrac{\slashed{P}_1 -\slashed{P}_3}{t^2 +i \epsilon} \gm^\nu +  \gm^\nu\dfrac{\slashed{P}_1 -\slashed{P}}{u^2 +i \epsilon} \gm^\mu $ and we have used total four momentum conservation. The squared amplitude is obtained by summing over colour, flavour and spin and is given  by
\begin{align}
	\sum \MB{\mM^{R,q\bar{q}\to g \gm }_{\rm tot}}^2 = \dfrac{1}{2} \dfrac{5\pi^2}{9}2^9 \alpha \alpha_S \FB{ \dfrac{t}{u} +\dfrac{u}{t}   }\label{Eq_M2_R_anni}~
\end{align}
where we have used $\Tr{t^a t^b} = \delta^{ab}/2$. 
In a similar way the squared matrix element for the annihilation of a left-handed quark and anti-quark pair is
\begin{align}
	\sum \MB{\mM^{L,q\bar{q}\to g \gm }_{\rm tot}}^2 = \dfrac{1}{2} \dfrac{5\pi^2}{9}2^9 \alpha \alpha_S \FB{ \dfrac{t}{u} +\dfrac{u}{t}   }~.\label{Eq_M2_L_anni}
\end{align}
The equality of Eqs.~\eqref{Eq_M2_R_anni} and \eqref{Eq_M2_L_anni} follows from the fact that the QED and QCD gauge interactions relevant to the present processes are vector-like and therefore couple identically to left- and right-handed quarks. Consequently, the corresponding invariant amplitudes are independent of chirality. Nevertheless, as will be shown later, the contributions of left- and right-handed quarks to the real photon production rate are generally different in the presence of a finite chiral chemical potential, $\mu_5$, since the corresponding distribution functions are modified by the chirality imbalance. The remaining contribution to the hard part of the photon production rate arises from QCD Compton scattering. The invariant amplitude for scattering of a right-handed quark with gluon as shown in Figs.~\ref{Fig_hard} (c) and (d) are given  by
\begin{align}
	\mM^{R,qg\to q\gm}_t &= -e g e_q \SB{\bar{u}_R (P_3)\gm^\mu t^a \dfrac{\slashed{P}_2 -\slashed{P}}{t^2 +i \epsilon} \gm^\nu u_R (P_2)  } \epsilon_\mu (P_1) \epsilon_\nu (P)  \\
	 \mM^{R,qg\to q\gm}_s &= -e g e_q \SB{\bar{u}_R (P_3)\gm^\nu t^a \dfrac{\slashed{P}_1 +\slashed{P}_2}{s^2 +i \epsilon} \gm^\mu u_R (P_2)  } \epsilon_\mu (P_1) \epsilon_\nu (P) ~.
\end{align}
Summing up the two contributions we get
\begin{align}
\mM^{R,qg\to q\gm}_{\rm tot} &=  -e g e_q {\bar{u}_R (P_3) t^a \Gamma_{R,qg\to q\gm}^\munu u_R (P_2)  } \epsilon_\mu(P_1) \epsilon_\nu (P)
\end{align}
with $\Gamma_{R,qg\to q\gm}^\munu   = \gm^\nu \dfrac{\slashed{P}_1 +\slashed{P}_2}{s^2 +i \epsilon} \gm^\mu- \gm^\mu \dfrac{\slashed{P}_1 -\slashed{P}_3}{t^2 +i \epsilon} \gm^\nu $. The squared amplitude is given by
 \begin{equation}
 	\sum \MB{\mM^{R,qg\to q\gm}_{\rm tot}}^2 =\frac{1}{2}  \dfrac{5\pi^2}{9}2^9 \alpha \alpha_S \FB{ \dfrac{t}{s} +\dfrac{s}{t}   }~\label{Eq_M2_R_comp}
 \end{equation}
In the similar way one can also calculate squared amplitude for QCD compton process for a left-handed quark with gluon which leads to the same result  as previously discussed :
 \begin{equation}
	\sum \MB{\mM^{L,qg\to q\gm}_{\rm tot}}^2 =\frac{1}{2}  \dfrac{5\pi^2}{9}2^9 \alpha \alpha_S \FB{ \dfrac{t}{s} +\dfrac{s}{t}   }~\label{Eq_M2_L_comp}.
\end{equation}
Figs.~\ref{Fig_hard} (e) and (f) corresponds to the QCD Compton process for a right-handed antiquark with gluon. Additionally one should also consider similar process for left-handed antiquarks. However since the coupling between quarks/anti-quarks  and gauge fields (both abelian and non-abelian) is invariant under charge conjugation the results will be same as Eqs.~\eqref{Eq_M2_R_comp} and \eqref{Eq_M2_L_comp} respectively. Now that we have the squared amplitudes for all three possible processes (annihilation and two Compton processes for  both left- and right-handed  quark and antiquark)  we can obtain the hard part of the photon production rate. Note that, in each diagram, all the particles taking part in the scattering process are thermalized except the outgoing photon due to its large mean free path as it does not take part in strong interaction.  The contribution to the differential rate from any of these processes is given by~\cite{Kapusta:1991qp,Baier:1991em} 
\begin{align}
	2E\frac{dR^{\rm \ hard}}{d^3 p} = \dfrac{1}{(2\pi)^8} \int \dfrac{d^3 \vec{p}_1}{2 E_1} \dfrac{d^3 \vec{p}_2}{2 E_2} \dfrac{d^3 \vec{p}_3}{2 E_3} \delta^{(4)} \FB{P_1 + P_2 -P_3 - P} f_1(E_1)f_2(E_2) \FB{1\pm f_3(E_3)} \sum\MB{\mM_{\rm tot}^{ \rm process}}^2 \label{Eq_rate_hard1} 
\end{align}
where $f$'s are Fermi-Dirac distribution functions for left- or right-handed fermions or Bose-Einstein distribution functions for gluons depending upon the particular process considered. The factor$\FB{1\pm f_3(E_3)}$ for the strongly interacting particle in the final state is either a Bose-enhancement with $(+)$ve sign applicable for annihilation processes or a Pauli-suppression with $(-)$ve sign applicable for two QCD Compton processes. Eq.~\eqref{Eq_rate_hard1} contains $9$-dimensional integrals and can be reduced to $4$-dimensional form assuming the outgoing photon is along $z$-direction (see Appendix~\ref{App_PhaseSpace} for detail). This yields
\begin{align}
		2E\frac{dR^{\rm \ hard}}{d^3 p} = \dfrac{1}{8 E (2\pi)^7} \int_{2 k_c^2}^{\infty}ds \int_{-s +k_c^2}^{-k_c^2} dt  \sum\MB{\mM_{\rm tot}^{ \rm process}}^2 \int_{\mathds{R}^2} dE_1 dE_2 \dfrac{\Theta\FB{\Xi (E_1,E_2)}  f_1 f_2 \FB{1\pm f_3} }{\sqrt{\Xi\FB{E_1,E_2}}} \label{Eq_rate_hard2}
\end{align}
where
\begin{align}
	\Xi \FB{E_1,E_2} &= -\FB{t E_1 +(s+t)E_2}^2 + 2 s E \FB{(s+t)E_2 - tE_1}-s^2 E^2 +s^2 t +st^2 \nn \\
	&= - (s+t)^2 E_2^2 + 2 (s+t) (sE-t E1)+ st (s+t) - (sE + t E_1)
\end{align}
and $\Theta $ is the step function. $k_c$ is an infrared cut-off to regulate the divergence of the invariant amplitudes  as $t$ and/or $u$ goes to zero. 
In the following we note down the distribution functions which appear for different cases 
\begin{align}
	f^{R,\pm}_{FD}(E) &= \dfrac{1}{\exp\FB{\dfrac{E\mp (\mu-\mu_5)}{T}}  +1},  \qquad\qquad 	f^{L,\pm}_{FD} (E) = \dfrac{1}{\exp\FB{\dfrac{E\mp (\mu+\mu_5)}{T}}  +1},  \\ \qquad &   \qquad\qquad\qquad f_{BE} (E) =\dfrac{1}{\exp\FB{\dfrac{E}{T}}  -1}~\label{Eq_dist_fns}.
\end{align}

For hard photon production, the photon energy satisfies $ E \gg T$. In this regime, one can employ the Boltzmann (classical) approximation and replace $f_1$ and $f_2$ by Boltzmann distribution functions instead of Fermi-Dirac or Bose-Einstein distribution functions so that $f_1 (E_1)f_2 (E_2) \approx e^{-(E_1+E_2)/T}$. At $\mu = \mu_5 = 0$, this allows one to perform all the integrals analytically, resulting in closed-form expressions for the differential rate of hard photon production via QCD Compton and annihilation processes, as shown in Refs.~\cite{Kapusta:1991qp,Baier:1991em,Haque:2024gva}. However, at finite values of $\mu$ (or $\mu_5$), the analytical calculation assuming the Boltzmann approximation as discussed in Ref.~\cite{Dumitru:1993us}, contains a term $\sim {\rm Ei} \FB{(4 \mu E - k_c^2)/4 E T}$ which encounters a branch point as  $\mu \to 0$. This leads to the following issues: (i) The established rate at $\mu\to 0$ is not recovered, and  (ii) analytical cancellation of the arbitrary cutoff parameter $k_c$ on addition of the soft and hard parts is not achieved. 
A similar problem appears even in case of nonzero $\mu_5$ as shown in Appendix~\ref{App_B}.
Therefore, following Ref.~\cite{Traxler:1994hy}, we evaluate the hard part given by Eq.~\eqref{Eq_rate_hard2} numerically for different physical processes using the exact Fermi-Dirac and Bose-Einstein distribution functions. Adding all the hard contributions to the soft contribution given by Eq.~\eqref{Eq_Soft_cont}, we now examine the independence of the final results on the arbitrary cut-off parameter $k_c$.

\section{Cut-off independence of the total photon rate}
\label{Sec_cutoff}

\begin{figure}[h]
	\includegraphics[scale=0.4]{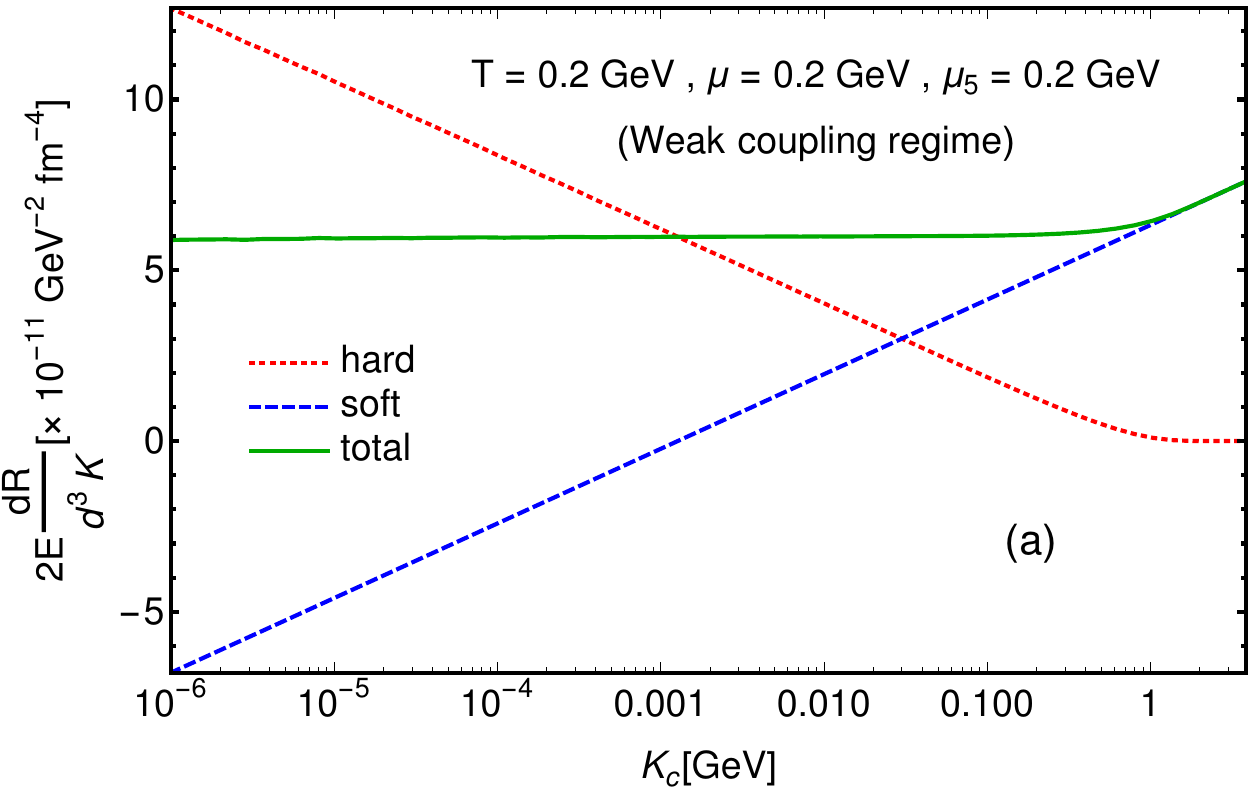}~~~
	\includegraphics[scale=0.4]{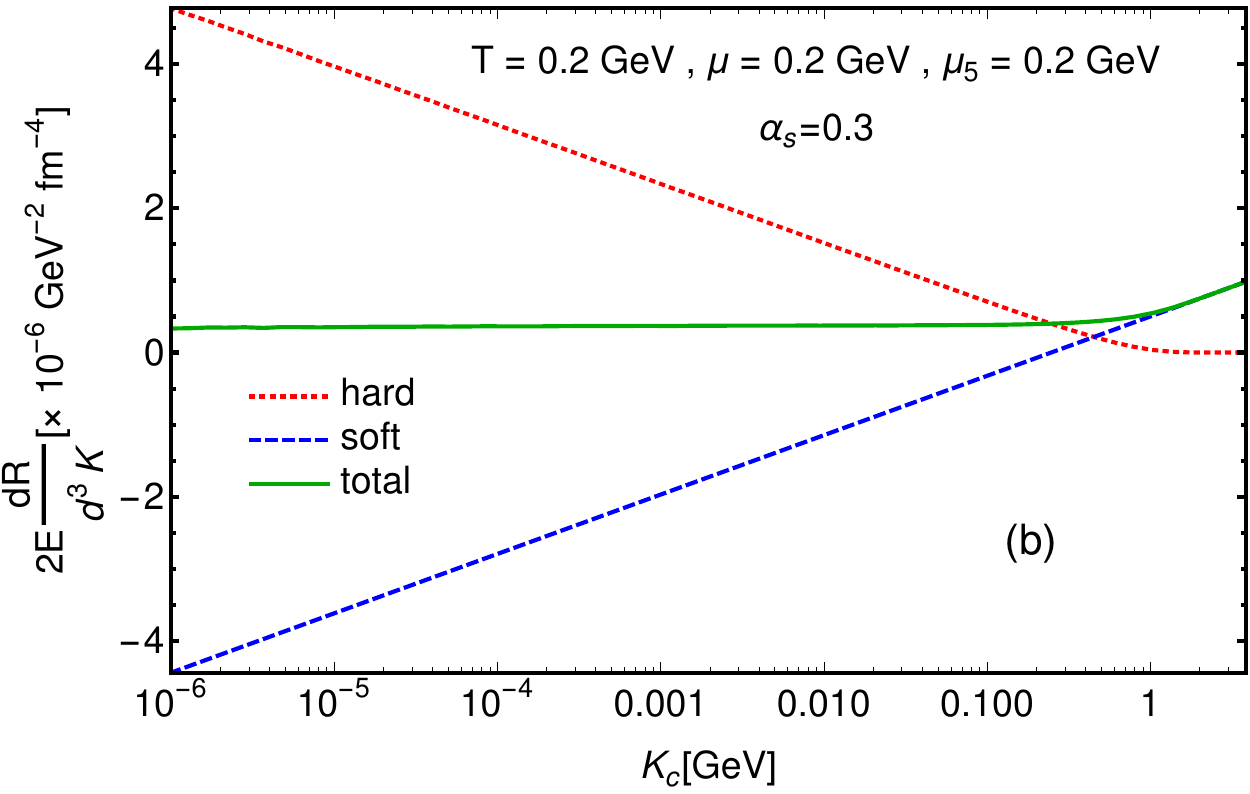}
	\caption{Total photon rate as function of arbitrary cut-off parameter $k_c$ is shown, along with the individual hard and soft contributions. The calculations are performed for $T = 200 $ MeV, $\mu = 200$ MeV and $\mu_5 = 200$ MeV. (a) Weak coupling regime with $g = 0.01$ (b) realistic scenario with $\alpha_S = 0.3$ which corresponds to $g \approx 1.94$. }
	\label{Fig_Cutoff}
\end{figure}

In Fig.~\ref{Fig_Cutoff}, we show the total photon production rate as a function of the arbitrary separation scale \(k_c\). The individual hard and soft contributions are also displayed separately. The cutoff independence of the total rate at finite baryon density has previously been established in Ref.~\cite{Traxler:1994hy}. Here, we extend this analysis to the case of finite chiral imbalance by considering \(\mu_5 = 200\) MeV together with \(T = 200\) MeV and \(\mu = 200\) MeV. In both panels, \(k_c\) is varied over nearly six orders of magnitude.
Since the cancellation of the \(k_c\)-dependence is formally expected in the weak-coupling limit, Fig.~\ref{Fig_Cutoff}(a) corresponds to \(g = 0.01\). The total photon rate remains essentially independent of \(k_c\) throughout the entire range considered, despite substantial variations in the individual hard and soft contributions. In particular, the soft contribution exhibits strong \(k_c\)-dependence and can become negative in certain regions, although the total rate remains positive and cutoff independent. Remarkably, the numerical cancellation between the hard and soft contributions remains highly efficient even in regions where the hierarchy \(gT \ll k_c \ll T\) is not strictly satisfied, and persists up to \(k_c \sim T\). This suggests that the Braaten--Pisarski resummation scheme remains valid in the presence of a finite chiral chemical potential.
In Fig.~\ref{Fig_Cutoff}(b), we present the corresponding result for the phenomenologically relevant value of \(\alpha_s = 0.3\), which corresponds to \(g \simeq 1.94\). The total photon rate again exhibits negligible dependence on \(k_c\) over the full range considered, indicating that the weak-coupling result remains numerically stable even for realistic values of the coupling constant. This observation is consistent with Refs.~\cite{Thoma:1993ii,Thoma:1993vs}, where it is argued that for quantities exhibiting logarithmic infrared divergences, the Braaten--Pisarski framework can remain applicable even at relatively large coupling provided the relevant particle energies are much larger than the temperature. Since the present analysis concerns hard photon production with energies significantly larger than the temperature scale, the extrapolation to phenomenologically relevant coupling is expected to remain reasonable.
The main difference between Figs.~\ref{Fig_Cutoff} (a) and (b) is that, for phenomenologically relevant values of the coupling constant, the range of \(k_c\) over which the total photon rate remains cutoff independent (while both the hard and soft contributions remain individually positive) is considerably narrower than in the weak-coupling limit.

\section{Total photon rate from chirally imbalanced QGP}
\label{Sec_result}
\begin{figure}[h]
	\includegraphics[scale=0.4]{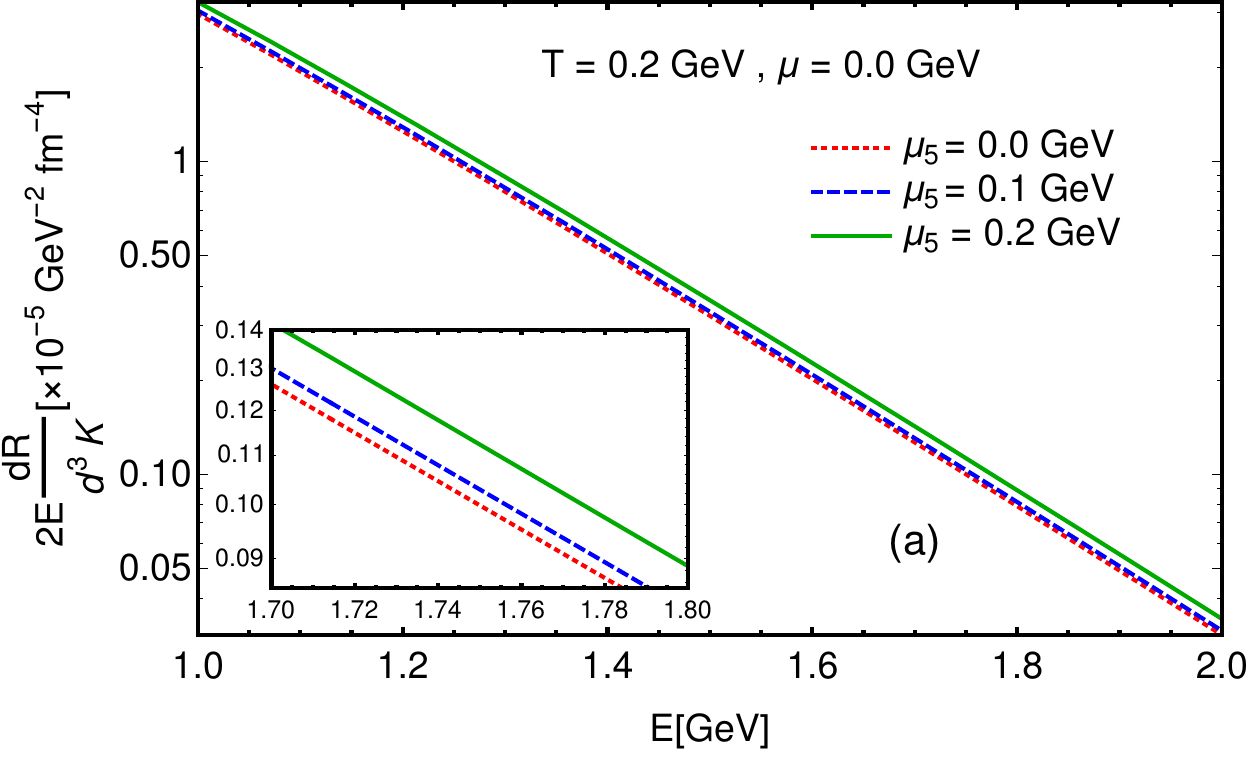}~~~
	\includegraphics[scale=0.4]{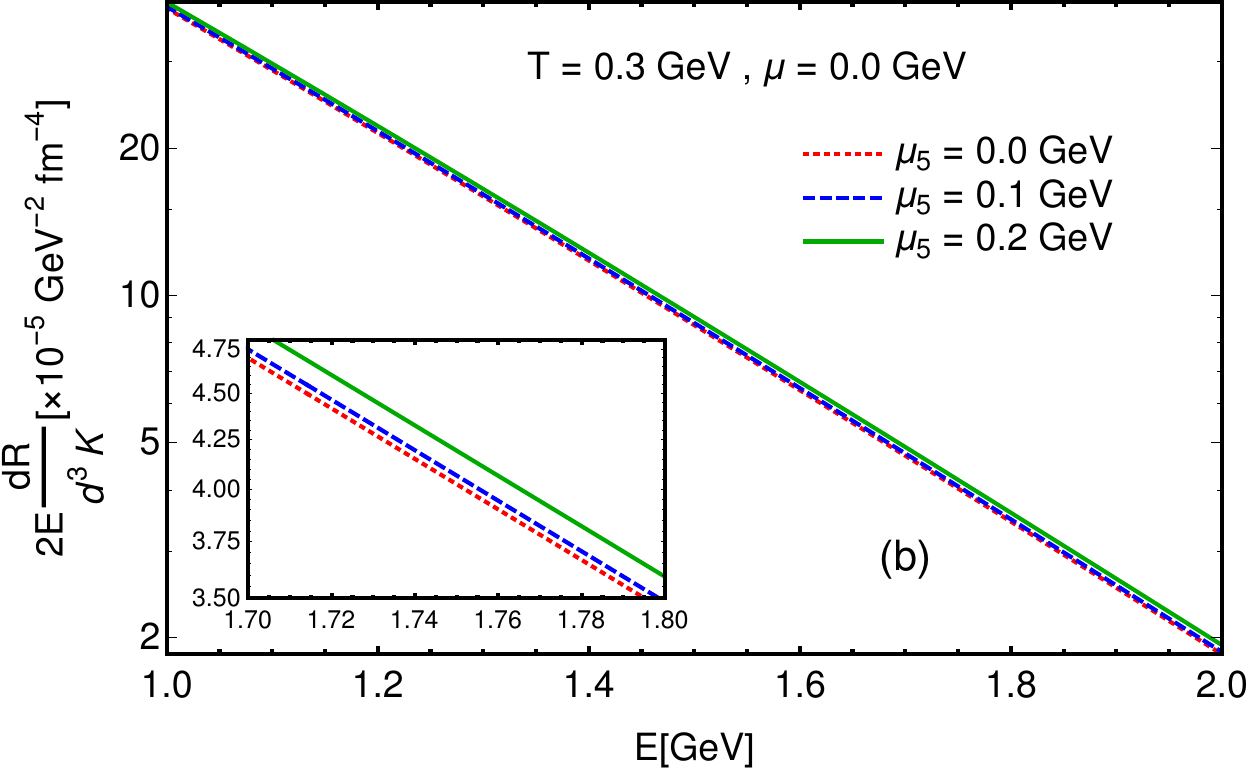}
	\caption{Photon production rate as a function of photon energy at vanishing baryon density. Plots are shown for  different values of \(\mu_5\) at (a) \(T = 200~\mathrm{MeV}\) and (b)  \(T = 300~\mathrm{MeV}\). All results are obtained for $\alpha_S = 0.3$.}
	\label{Fig_ERate}
\end{figure}
As discussed in the Sec.~\ref{Sec_Intro}, the systems formed in heavy-ion collisions may exhibit a net chirality with random sign for domains that are preferentially right- or left-handed. In addition, the net baryon density in the produced matter is expected to be negligible at collider energies. Motivated by this, we set \(\mu = 0\) throughout this section. This choice isolates the effects of chiral imbalance on the observables under consideration.
Without any loss of generality, we further restrict the analysis to a left-handed configuration by taking \(\mu_5 > 0\). This convention is adopted solely for definiteness and does not affect the generality of the results.
Following the discussions of Sec.~\ref{Sec_cutoff} we will choose $k_c = 0.5 T$ so that the total photon rate remains independent of the cutoff parameter at realistic values of $\alpha_S$.

In Figs~\ref{Fig_ERate} (a) and (b) we have shown the variation of total production rate of high energy photon at (a) \(T = 200~\mathrm{MeV}\) and (b)  \(T = 300~\mathrm{MeV}\) respectively for different values of $\mu_5$ at vanishing baryon density. In both cases, the photon spectra exhibit a pronounced exponential suppression with increasing photon energy, indicating the thermal origin of the emission process and reflecting the diminishing probability of populating high-energy states in the medium.

 It can be seen that the spectra are dominated by the exponential decrease with photon energy. From both the figures it is evident that the presence of chiral imbalance results in an additional enhancement in the rate. A comparison between Figs.~\ref{Fig_ERate}(a) and (b) shows that the production rate increases with temperature, reflecting the increase of available thermal phase space. A notable feature emerging from both panels is the enhancement of the photon production rate in the presence of chiral imbalance. As the value of $\mu_5$ increases, the rate systematically rises over the entire energy range considered as shown in the inset plot. This enhancement originates from the modification of the underlying quark and antiquark distribution functions induced by the chiral asymmetry, which alters the available phase space and scattering probabilities contributing to photon emission. Consequently, a finite chiral chemical potential increases the abundance of effective number of participants in the plasma leading to a larger net photon yield. A comparison between Figs.~\ref{Fig_ERate}(a) and (b) further reveals a substantial increase in the photon production rate with temperature. This enhancement can be attributed to the growth of the thermal phase space as higher temperatures increase the occupation of energetic quark and antiquark states participating in photon-producing processes. Moreover, thermal excitations become more abundant, thereby increasing the probability of scattering and annihilation processes responsible for hard photon emission.

\begin{figure}[h]
	\includegraphics[scale=0.4]{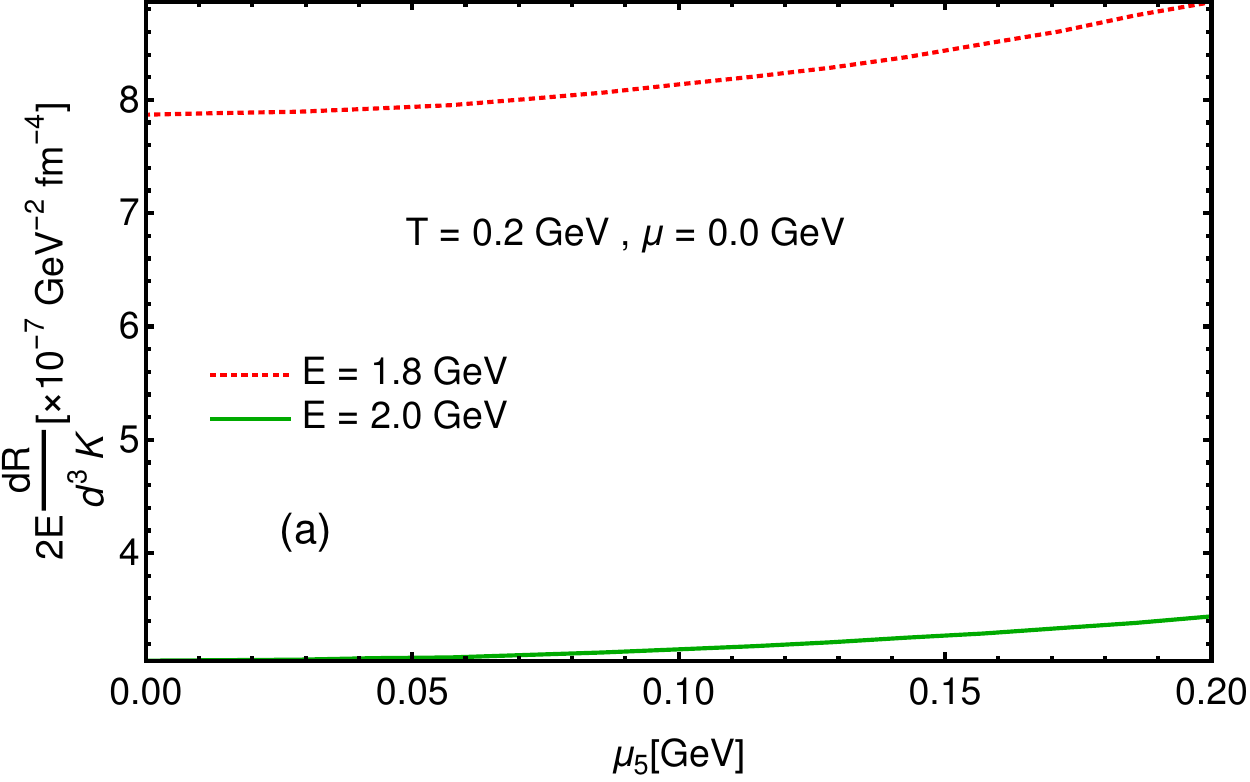}~~~
	\includegraphics[scale=0.4]{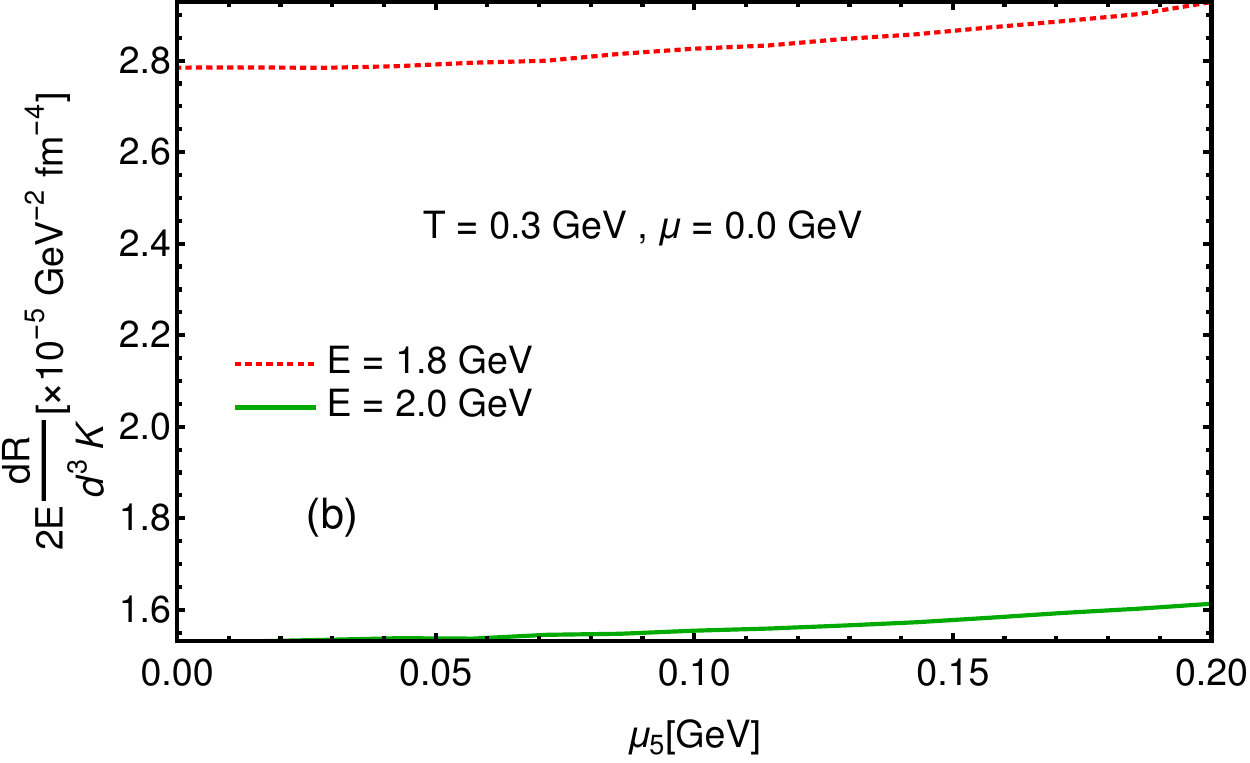}
	\caption{Photon production rate as a function of \(\mu_5\) at vanishing baryon density. Plots are shown for different photon energies at (a) \(T = 200~\mathrm{MeV}\) and (b) \(T = 300~\mathrm{MeV}\). All results are obtained using \(\alpha_s = 0.3\). }
	\label{Fig_mu5Rate}
\end{figure}
We now study the photon production rate as a function of \(\mu_5\) for different temperatures. In Fig.~\ref{Fig_mu5Rate}(a), we show the \(\mu_5\)-dependence of hard photon production at \(T = 200~\mathrm{MeV}\) for photon energies \(E = 1.8\) and \(2~\mathrm{GeV}\). The rate increases with increasing \(\mu_5\), and the enhancement is more pronounced at lower photon energy.
A qualitatively similar behaviour is observed in Fig.~\ref{Fig_mu5Rate}(b) for \(T = 300~\mathrm{MeV}\), with all other parameters held fixed. A comparison between Figs.~\ref{Fig_mu5Rate}(a) and (b) shows that the production rate increases with temperature, reflecting the increase of available thermal phase space as previously explained.

\section{Summary \& outlook}
\label{Sec_summary}
In this work we calculate the thermal photon emission rate from a quark gluon plasma with chiral imbalance. The hard part is obtained using matrix elements for annihilation and QCD Compton processes involving left- and right-handed quarks. These involve an arbitrary momentum cutoff  parameter $k_c$ introduced in order to tame the infrared divergences. The $k_c$-dependence is cancelled by adding the soft contribution obtained by us in a previous work~\cite{Chaudhuri:2025kpl}. The cutoff independence of the total rate at finite values of chiral chemical potential is demonstrated in the weak coupling limit as well as for phenomenologically relevant values of strong coupling constant.  The numerical cancellation of the separation scale between the hard and soft contributions remains efficient for a wide range of values. The presence of chiral imbalance is observed to cause an overall enhancement in the emission rate of thermal photons. This can be attributed to additional terms proportional to the square of the chiral chemical potentials in the Debye mass as well as to the increase in abundance of effective number of participants in the chiral plasma. In order to estimate the chiral chemical potential by making a connection with the experimentally observed photon spectra this rate has to be convoluted with the space time evolution using relativistic hydrodynamics with axial anomaly~\cite{Guo:2017jxs}.  

\appendix
\section{Reduction of the phase-space integral}
\label{App_PhaseSpace}

In this appendix, we provide the detailed derivation of Eq.~\eqref{Eq_rate_hard2} from Eq.~\eqref{Eq_rate_hard1} following Refs~\cite{Kapusta:1991qp,Haque:2024gva}. 
The differential photon production rate from any $2\to 2$ scattering process is given by,
\begin{align}
	2E\frac{dR^{\rm \ hard}}{d^3 p} 
	= \dfrac{1}{(2\pi)^8} \int \dfrac{d^3 \vec{p}_1}{2 E_1} \dfrac{d^3 \vec{p}_2}{2 E_2} \dfrac{d^3 \vec{p}_3}{2 E_3} \delta^{(4)} \FB{P_1 + P_2 -P_3 - P} f_1(E_1)f_2(E_2) \FB{1\pm f_3(E_3)} \sum\MB{\mM_{\rm tot}^{ \rm process}}^2 , \tag{14}
\end{align}
where $P=(E,\vec{p})$ is the four-momentum of the outgoing photon. The Mandelstam variables for the processes shown in Fig.~\ref{Fig_hard} are given by,
\[
s = (P_1 + P_2)^2, \qquad t =(P_2-P)^2 = (P_1 - P_3)^2, \qquad u = (P_2-P_3)^2 = (P_1 - P)^2,
\]
which for massless particles satisfy $s+t+u=0$. Inserting unity in the form of integral representations of Dirac delta functions for $s$ and $t$,
\[
1 = \int ds \, \delta[s - (P_1 + P_2)^2] \int dt \, \delta[t - (P_2 - P)^2],
\]
the rate becomes
\begin{align}
	2E\frac{dR^{\rm \ hard}}{d^3 p} 
	&= \dfrac{1}{(2\pi)^8} \int ds \, dt \int \dfrac{d^3 \vec{p}_1}{2 E_1} \dfrac{d^3 \vec{p}_2}{2 E_2} \dfrac{d^3 \vec{p}_3}{2 E_3} f_1 f_2 \FB{1\pm f_3} \sum\MB{\mM(s,t)}^2 \nonumber \\
	&\quad \times \delta^{(4)}(P_1 + P_2 -P_3 - P) \delta[s - (P_1 + P_2)^2] \delta[t - (P_2 - P)^2].
\end{align}
Now we make the replacement $ \int \frac{d^3 \vec{p}_3}{2 E_3}  \to \int d^4 P_3 \Theta (p^0_3) \delta (P^2) $ and integrate over $P_3$ using $ \delta^{(4)}(P_1 + P_2 -P_3 - P)$ which yields
\begin{align}
	2E\frac{dR^{\rm \ hard}}{d^3 p} 
	&= \dfrac{1}{(2\pi)^8} \int ds \, dt \int \dfrac{d^3 \vec{p}_1}{2 E_1} \dfrac{d^3 \vec{p}_2}{2 E_2} f_1 f_2 \FB{1\pm f_3} \sum\MB{\mM(s,t)}^2 \nonumber \\
	&\quad \times \delta[t - (P_2 - P)^2] ~~ \delta[(P_1 + P_2 - P)^2] ~~\delta[s - (P_1 + P_2)^2] ~\Theta(E_1 + E_2 - E) \label{Eq_app_rate3}.
\end{align}
At this stage, the phase space is still 6-dimensional (integrations over $\vec{p}_1$ and $\vec{p}_2$). We choose the photon momentum to lie along the $z$-axis so that:
\[
P = E(1, 0, 0, 1),
\]
and the momenta of the incoming particles in spherical coordinates are:
\begin{align}
	P_1 &= E_1(1, \sin\theta_1 \cos\phi_1, \sin\theta_1 \sin\phi_1, \cos\theta_1),\label{Eq_App_angle_conv1}\\
	P_2 &= E_2(1, \sin\theta_2 \cos\phi_2, \sin\theta_2 \sin\phi_2, \cos\theta_2)~.\label{Eq_App_angle_conv2}
\end{align}
Using the arguments of the $\delta$-function in Eq.~\eqref{Eq_app_rate3} we obtain the following relations
\begin{align}
	\cos\theta_2 &= 1 + \frac{t}{2E E_2},\\
	\cos\theta_1 &= \frac{2EE_1 - s - t}{2EE_1},\\
	\cos(\phi_1 - \phi_2) &= \frac{-2E^2 s - 2EE_1 t + (s+t)(2EE_2 + t)}{4E^2 E_1 E_2 \sin\theta_1 \sin\theta_2}. \label{Eq_App_cosphi12}
\end{align}
Note that using the parametrization given in Eqs.~\eqref{Eq_App_angle_conv1} and \eqref{Eq_App_angle_conv2} the phase-space measure is given by
\[
\frac{d^3\vec{p}_1}{2E_1} \frac{d^3\vec{p}_2}{2E_2}
= \frac{1}{4} E_1 dE_1 \, d\cos\theta_1 \, d\phi_1 \, E_2 dE_2 \, d\cos\theta_2 \, d\phi_2.
\]
Now using the following relation 
\begin{equation}
	\delta (f(x)) = \sum_{x_0}\dfrac{\delta (x-x_0)}{\MB{f^\prime (x_0)}}.
\end{equation}
where $x_0$ are the roots of $f(x)$, one can evaluate the integrals over the angular variables. Doing the $\cos \theta_1 $ and $\cos \theta_2 $‑integrals using  $ \delta[t - (P_1 - P)^2]$ and $\delta[(P_1 + P_2 - P)^2]$ respectively we obtain
\begin{align}
	2E\frac{dR^{\rm \ hard}}{d^3 p} 
	&= \dfrac{1}{(2\pi)^8} \int ds \, dt \int \dfrac{E_1 dE_1}{2} \TB{\dfrac{1}{2 E E_1}}\int \dfrac{E_2 dE_2}{2} \TB{\dfrac{1}{2 E E_2}}  f_1 f_2 \FB{1\pm f_3} \sum\MB{\mM(s,t)}^2 \nonumber \\
	&\quad \times \int d \phi_1 ~ d \phi_2 ~~\delta[\Phi] ~\Theta(E_1 + E_2 - E) \label{Eq_app_rate4}.
\end{align}
Here the two bracketed factors ${\frac{1}{2E E_1}}$ and  ${\frac{1}{2E E_2}}$ are the Jacobians coming from integrations over $\theta_1$ and $\theta_2$ respectively. What remains of $\delta[s - (P_1 + P_2)^2]$ after $\theta_1$ and $\theta_2$ are fixed is denoted as $\delta(\Phi)$. First the $\phi_1$ integration is done using $\delta$-function and then integrating over $\phi_2$  one gets
\begin{align}
	2E\frac{dR^{\rm \ hard}}{d^3 p} 
	&= \dfrac{1}{(2\pi)^7}\dfrac{1}{8 E} \int ds \, dt \int d E_1 dE_2~  f_1 f_2 \FB{1\pm f_3} \sum\MB{\mM(s,t)}^2   ~\dfrac{\Theta(\Xi(E_1,E_2))}{\sqrt{\Xi(E_1,E_2)}} \label{Eq_app_rate5}
\end{align}
where $\Xi(E_1,E_2) = -\FB{t E_1 +(s+t)E_2}^2 + 2 s E \FB{(s+t)E_2 - tE_1}-s^2 E^2 +s^2 t +st^2 $. An extra factor of $2$ comes from the fact that there are two values of $\phi_1$ that satisfies $\cos (\phi_1 - \phi_2) $ given by Eq.~\eqref{Eq_App_cosphi12}. The Jacobian coming from integration over $\phi_1$ is given by $\frac{1}{2 E1 E2 \sin \theta_1 \sin \theta_2 \sin (\phi_1-\phi_2)} = \frac{E}{\sqrt{\Xi(E_1,E_2)}}$. The function $\Theta(\Xi(E_1,E_2))$ appears from the condition $ \MB{\cos (\phi_1 - \phi_2) }\le 1$. Thus we arrive at Eq.~\eqref{Eq_rate_hard2}.

\section{Analytical calculation for cutoff independence}
\label{App_B}
\subsection{Analytical calculation of the hard part with Boltzmann approximation}
To carry out analytical calculation starting from Eq.~\eqref{Eq_rate_hard2}, let us introduce $V = E_1 + E_2$ so that 
\[
\int dE_1 \, dE_2 \longrightarrow \int dV \, dE_2
\]
Replacing $E_1$ by $E_1 = V - E_2$, we we can write 
\[
\Xi(E_2, V) = \left( A E_2^2 + B E_2 + C \right)^{1/2}
\]
where
\begin{align}
	A &= -s^2 \qquad
	B = 2s \left( sE + 2tE - tV \right) \qquad
	C = st (s + t) - (sE + tV)^2~.
\end{align}
Limits of these integrations over $E_2$ and $V$ are fixed by delta functions in Appendix~\ref{App_PhaseSpace} of the present MS and can be estimated as
\begin{align}
	-1 &\le \cos\theta_1 \le 1
	\quad\Rightarrow \quad (s + t) \ge 0 \quad \text{and} \quad E_1 \ge \frac{s + t}{4E} \\[1ex]
	-1 &\le \cos\theta_2 \le 1 \quad
	\Rightarrow \quad t < 0 \quad \text{and} \quad E_2 \ge -\frac{t}{4E}\\
	-1 &\le \cos(\phi_1 -\phi_2)\le 1 \quad \Rightarrow \quad A E_2^2 + B E_2 + C \ge 0	
\end{align}
Moreover one can write $ \sqrt{ A E_2^2 + B E_2 + C}= s\sqrt{(E_2^{\rm max}-E_2) (E_2- E_2^{\rm min})}$ and the condition $B^2-4AC \ge 0$ leads to 
\[
4 s^2 (-t) (s + t) \left( s - 4EV + 4E^2 \right) \ge 0 \quad \Rightarrow \quad V \ge \frac{s}{4E} + E
\]
Using the above expressions we can write Eq.~\eqref{Eq_rate_hard2} as
\begin{align}
	2E\frac{dR^{\rm \ hard}}{d^3 p} 
	&= \dfrac{1}{(2\pi)^7}\dfrac{1}{8 E} \int \dfrac{ds}{s} \, dt \int_{s/4E +E}^{\infty} d V   \sum\MB{\mM(s,t)}^2 \nn \\
	&~~~~~~~~~~~\times  \int_{E_2^{\rm min}}^{E_2^{\rm max}} dE_2~   \dfrac{1}{\sqrt{(E_2^{\rm max}-E_2) (E_2- E_2^{\rm min})}} \sum\MB{\mM(s,t)}^2 ~f_1 f_2 \FB{1\pm f_3} 
	\label{Eq_rate_hard3}.
\end{align}
 Up to this point the result is exact and now we will introduce some physically motivated approximations to obtain analytical result. Using the fact that the photon energy is hard we will replace the distribution functions of incoming particles by Maxwell-Boltzmann distribution function. Since we are dealing with non-zero $\mu$ and $\mu_5$ we have to treat the annihilation and Compton processes separately. 
\subsubsection*{Pair annihilation process}
Let us consider the process $q_R \ov{q}_R \to g \gm $. In this case we have
\begin{align}
	f_1 f_2 &= f_{\text{FD}}^{R,+} (E_1) \, f_{\text{FD}}^{R,-} (E_2) \simeq e^{-(E_1 - \mu_R)/T} \, e^{-(E_2 + \mu_R)/T} = e^{-V/T} \nn \\
	f_3 &= f_{\text{BE}}(E_3) = f_{\text{BE}}(V - E)
\end{align}
So, in this case after taking MB approximation effect of $\mu_R$ disappears. The integral over $E_2$ can be performed analytically yielding a factor of $\pi$. Similar situation will also appear for $q_L \ov{q}_L \rightarrow g \gamma$ scenario. So, following previous calculations~\cite{Kapusta:1991qp,Baier:1991em,Haque:2024gva} one can write:
\begin{align}
	2E \frac{dR^{q_R \ov{q}_R \rightarrow g \gamma}}{d^3 p} = 2E \frac{dR^{q_L \ov{q}_L \rightarrow g \gamma}}{d^3 p} = \frac{5\alpha\alpha_s}{27\pi^2} \, T^2 e^{-E/T} \left[ \ln \frac{4ET}{k_c^2} - 1 - \gamma_E + \frac{\zeta'(2)}{\zeta(2)} \right] \label{Eq_Anni_RL}
\end{align}

\subsubsection*{Compton scattering}

Let us first consider the process $q_R \, g \rightarrow q_R \, \gamma$. In this case we get
\begin{align*}
	f_1 f_2 &= f_{\text{FD}}^{R,+} (E_1) \, f_{\text{BE}}(E_2) \simeq e^{-(E_1 + E_2 - \mu_R)/T} = e^{-(V - \mu_R)/T}
\end{align*}
Now we perform $V$-integral in Eq.~\eqref{Eq_rate_hard3} to obtain
\begin{align*}
	&\int_{\frac{s}{4E} + E}^{\infty} dV \, e^{-(V - \mu_R)/T} \left( 1 - \frac{1}{e^{(V - E - \mu_R)/T} + 1} \right) \nn \\
	&~~~~= T e^{-E/T} \int_{\frac{s}{4ET} - \frac{\mu_R}{T}}^{\infty} \frac{dz}{e^z + 1} = T e^{-E/T} \ln \left[ 1 + \exp \FB{-\dfrac{s - 4\mu_R E}{4ET}} \right]
\end{align*}
where $z = \frac{V - E + \mu_R}{T} $. Using Eq.~\eqref{Eq_M2_R_comp} of the present MS we arrive at
\begin{align}
	2E \frac{dR^{q_R g\rightarrow q_R \gamma}}{d^3 p} &= \dfrac{1}{(2\pi)^6}\dfrac{1}{16 E} \int \dfrac{ds}{s}   T e^{-E/T} \ln \TB{ 1 + \exp \FB{-\dfrac{s - 4\mu_R E}{4ET}} }\int_{-s+k_c^2}^{k_c^2} \, dt  \frac{1}{2}  \dfrac{5\pi^2}{9}2^9 \alpha \alpha_S \FB{ \dfrac{t}{s} +\dfrac{s}{t}   } \nn \\
	&= \dfrac{5 \alpha\alpha_S}{9 \pi^4} \dfrac{T e^{-E/T}}{8E} \int_{2 k_c^2}^\infty \dfrac{ds}{s}  \ln \TB{ 1 + \exp \FB{-\dfrac{s - 4\mu_R E}{4ET}} } \TB{s - 2k_c^2 + 2 s \ln \FB{\dfrac{s-k_c^2}{k_c^2}}}
\end{align}
The $s-$integral appearing in the above expression is same as that given in~\cite{Dumitru:1993us} upon replacing ($\mu_R \to\mu$) and this will lead to the term $\sim {\rm Ei} \FB{(4 \mu_R E - k_c^2)/4 E T}$ which encounters a branch point as $\mu_R \to 0$. Thus we arrive at the same problem as in case of finite quark chemical potential as discussed in~\cite{Traxler:1994hy}.

\subsection{Small $\mu_5$ approximation}

Now  let us try to derive the same assuming the case $\mu=0$ and small values of $\mu_5$. In this case one can make the following expansion
\begin{align}
	f_{\text{FD}}^{R,\pm} (E,\mu_5) &=\dfrac{1}{e^{(E\pm\mu_5)/T}+1} \nn \\
	&= \dfrac{1}{e^{E/T} +1} \mp \dfrac{\mu_5}{T} \dfrac{e^{E/T}}{(e^{E/T} +1)^2}+ \half \FB{\dfrac{\mu_5}{T}}^2 \dfrac{e^{E/T} (e^{E/T}-1)}{(e^{E/T} +1)^3}	 + \mO \FB{\FB{\dfrac{\mu_5}{T}}^3} \nn \\
	& \simeq \FB{1 \mp \dfrac{\mu_5}{T} + \half \FB{\dfrac{\mu_5}{T}}^2} e^{-E/T}
\end{align} 
where in the last line we have assumed $E \gg T$ and kept terms up to $\mO\FB{\FB{\dfrac{\mu_5}{T}}^2} $. In a similar way one can write
\begin{align}
	f_{\text{FD}}^{L,\pm}(E,\mu_5) \simeq \FB{1 \pm \dfrac{\mu_5}{T} + \half \FB{\dfrac{\mu_5}{T}}^2} e^{-E/T}~.
\end{align}
Let us now concentrate on the different scattering processes separately.
\subsubsection*{Annihilation}
For the process $q_R \ov{q}_R \to g \gm $ we have
\begin{align}
	f_1 f_2 = f_{\text{FD}}^{R,+} (E_1) \, f_{\text{FD}}^{R,-} (E_2) &\simeq \FB{1 - \dfrac{\mu_5}{T} + \half \FB{\dfrac{\mu_5}{T}}^2} \FB{1 + \dfrac{\mu_5}{T} + \half \FB{\dfrac{\mu_5}{T}}^2} e^{-E_1 /T} \, e^{-E_2/T}\nn \\
	& = \FB{1+ \FB{\dfrac{\mu_5}{T}}^4} e^{-V/T} \simeq e^{-V/T}\nn \\
	f_3 &= f_{\text{BE}}(E_3) = f_{\text{BE}}(V - E)
\end{align}
where we have kept terms upto $\mO\FB{\FB{\dfrac{\mu_5}{T}}^2} $.  Thus here also we get the same result as given by Eq.~\eqref{Eq_Anni_RL}. Thus the annihilation rate adding the contributions from both left- and right-handed quarks becomes
\begin{align}
	2E \frac{dR^{\rm ~Annihilation}}{d^3 p}  = \frac{5\alpha\alpha_s}{27\pi^2} \, 2 T^2 e^{-E/T} \left[ \ln \frac{4ET}{k_c^2} - 1 - \gamma_E + \frac{\zeta'(2)}{\zeta(2)} \right] \label{Eq_Anni_tot}
\end{align}

\subsubsection*{Compton}

For the process $q_R \, g \rightarrow q_R \, \gamma$ using the same approach we get
\begin{align}
	f_1f_2
	&\simeq
	e^{-(E_1-\mu_5)/T}e^{-E_2/T}
	\simeq
	\left(
	1+\frac{\mu_5}{T}+\frac{\mu_5^2}{2T^2}
	\right)e^{-V/T},
	\\
	1-f_3
	&=
	1-\frac{1}{e^{(E_3-\mu_5)/T}+1}
	=
	1-\frac{1}{e^{(V-E-\mu_5)/T}+1}
	\simeq 1-\frac{1}{e^{(V-E)/T}+1}\left(
	1+\frac{\mu_5}{T}+\frac{\mu_5^2}{2T^2}
	\right)
\end{align}
Let us define $ \xi(\mu_5)\equiv 1+\frac{\mu_5}{T}+\frac{\mu_5^2}{2T^2}$ which leads to the following form of the $V$-integral starting from Eq.~\eqref{Eq_rate_hard3}
\begin{align}
	\int dV\,\xi(\mu_5)\,e^{-V/T}
	&\left(
	1-\frac{1}{e^{(V-E)/T}+1}\xi(\mu_5)
	\right)=\xi(\mu_5)\int dV\,e^{-V/T}
	-\xi(\mu_5)^2\int dV\,
	\frac{e^{-V/T}}
	{e^{(V-E)/T}+1}.
	\nn \\
	& = \xi(\mu_5) T e^{-E/T}e^{-s/(4ET)}
	-\xi(\mu_5)^2 T e^{-E/T}
	\int_{s/(4ET)}^\infty
	\frac{e^{-x}}{e^x+1}\,dx     \qquad\qquad \text{with~} ~x=\frac{V-E}{T}   \nn \\
	& = \xi(\mu_5) T e^{-E/T}e^{-s/(4ET)}
	-
	\xi^2(\mu_5)T e^{-E/T}
	\left\{
	e^{-s/(4ET)}
	+\frac{s}{4ET}
	-\ln\left[1+e^{s/(4ET)}\right]
	\right\}, \nn
	\\
	&=
	T e^{-E/T} \TB{
		\FB{
			-\frac{\mu_5}{T}
			-\frac32\frac{\mu_5^2}{T^2}
		} e^{-s/(4ET)}
		+
		\FB{
			1+\frac{2\mu_5}{T}
			+\frac{2\mu_5^2}{T^2}
		} ~ \ln\left[1+e^{-s/(4ET)}\right] }~.
\end{align}
The $t$-integration is trivial and is given by
\begin{align}
	\int_{-s+k_c^2}^{-k_c^2}dt
	\left(\frac{s}{t}+\frac{t}{s}\right)
	&=
	\frac{1}{2}\left\{s-2k_c^2+2s\ln\left(\frac{s-k_c^2}{k_c^2}\right)
	\right\},
\end{align}
Now let us note the following expressions
\begin{align}
	\int_0^\infty dy\,\ln(1+e^{-y}) &=\frac{\pi^2}{12} \\
	\int_0^\infty dy\,\ln y\,\ln(1+e^{-y})
	&=
	\sum_{n=1}^\infty
	\frac{(-1)^{n+1}}{n}
	\int_0^\infty dy\,e^{-ny}\ln y
	= \sum_{n=1}^\infty
	\frac{(-1)^{n+1}}{n}
	\left[
	\frac{-\gamma_E-\ln n}{n}
	\right]. \\
	\eta(s) = \sum_{n=1}^{\infty} \frac{(-1)^{n-1}}{n^s} &= (1 - 2^{1-s})\zeta(s) \\[1ex]
	\eta'(s) = \sum_{n=1}^{\infty} \frac{(-1)^n \ln n}{n^s} &= 2^{1-s} \ln 2 \, \zeta(s) + (1 - 2^{1-s})\zeta'(s)
\end{align}
where $\eta(s)$ is Dirichlet eta function and $\zeta (s)$ is Riemann zeta function. Using these identities one can perform the remaining integrals. After some algebra we arrive at
\begin{align}
	2E\frac{dR^{q_R g\rightarrow q_R \gamma}}{d^3p}
	&=
	\frac{5\alpha\alpha_s}{108\pi^2}
	T^2e^{-E/T}
	\Bigg[
	\frac{12}{\pi^2}
	\left(
	-\frac{\mu_5}{T}
	-\frac32\frac{\mu_5^2}{T^2}
	\right)
	\left(
	\frac12+\ln\frac{4ET}{k_c^2}-\gamma_E
	\right)
	\nonumber\\
	&\qquad+
	\left(
	1+\frac{2\mu_5}{T}
	+\frac{2\mu_5^2}{T^2}
	\right)
	\left(
	\frac12+\ln\frac{4ET}{k_c^2}
	-\gamma_E+\ln2+\frac{\zeta'(2)}{\zeta(2)}
	\right)
	\Bigg].
\end{align}
 The corresponding contribution for the process $q_L g \to q_L \gamma$ is obtained by replacing $\mu_5$ with $ - \mu_5$. The antiquark contribution is identical to the quark contribution in the absence of baryon chemical potential. Consequently, for small values of $\mu_5$, the total Compton contribution to the hard-photon production rate in a medium with vanishing baryon density is given by
\begin{align}
	2E\frac{dR^{\rm ~Compton}}{d^3p}
	&=
	\frac{5\alpha\alpha_s}{27\pi^2}
	T^2e^{-E/T}
	\Bigg[
	-\frac{18}{\pi^2}
	\frac{\mu_5^2}{T^2}
	\left(
	\frac12+\ln\frac{4ET}{k_c^2}-\gamma_E
	\right)
	\nonumber\\
	&\qquad+
	\left(
	1+
	\frac{2\mu_5^2}{T^2}
	\right)
	\left(
	\frac12+\ln\frac{4ET}{k_c^2}
	-\gamma_E+\ln2+\frac{\zeta'(2)}{\zeta(2)}
	\right)
	\Bigg].\label{Eq_Comp_tot}
\end{align}
Adding Eqs.~\eqref{Eq_Anni_tot} and \eqref{Eq_Comp_tot} we obtain the total hard contribution to the thermal photon production rate
\begin{align}
	2E \frac{dR^{\rm hard}}{d^3 p}
	&= \dfrac{5 \alpha\alpha_S}{9 \pi^2} {T^2 e^{-E/T}} \TB{ \ln \dfrac{4ET}{k_c^2}- \dfrac{1}{2} -\gm_E + \dfrac{1}{3}\ln 2 + \dfrac{\zeta^\prime(2)}{\zeta(2)} } \nn \\
	& ~~~~ + \dfrac{5 \alpha\alpha_S}{9 \pi^2} {\mu_5^2 e^{-E/T}} \dfrac{2 \zeta (2) - 3}{3 \zeta (2)} \TB{ \ln \dfrac{4ET}{k_c^2}+ \dfrac{1}{2} -\gm_E + \dfrac{2 \zeta (2)}{2 \zeta (2) - 3} \FB{\ln 2 + \dfrac{\zeta^\prime(2)}{\zeta(2)}} }   \label{Eq_hard_final}
\end{align}
The first term matches exactly with the previous result obtained in~\cite{Kapusta:1991qp,Baier:1991em,Haque:2024gva} in absence of $\mu_5$. The second term corresponds to non-trivial finite $\mu_5$-correction. Now following Eqs.~(2) and (3) of the present MS at $\mu=0$ we can write 
\begin{align}
	2E \frac{dR^{\rm soft}}{d^3 p} 
	&= \dfrac{5 \alpha\alpha_S}{9 \pi^2} {T^2 e^{-E/T}}  \ln \dfrac{k_c^2}{2 M^2} + \dfrac{5 \alpha\alpha_S}{9 \pi^2} \dfrac{\mu_5^2}{\pi^2} e^{-E/T}  \ln \dfrac{k_c^2}{2 M^2} \label{Eq_soft_final}
\end{align}
where $M^2 =  \dfrac{g^2  C_F}{8} (T^2 + \dfrac{\mu_5^2}{\pi^2})$. Comparing  Eqs.~\eqref{Eq_hard_final} and \eqref{Eq_soft_final} one can conclude that cancellation of the intermediate momentum cutoff parameter $k_c$ is not possible for the nonzero $\mu_5$ part.

\bibliography{reference} 

	\end{document}